\documentclass[useAMS,usenatbib]{mn2e}
\usepackage{epsfig}
\usepackage{times}
\usepackage{amssymb}
\usepackage{amsbsy}
\newcommand{\uv}{{\it uv }}

\title[Reionization with higher order statistics]{Detection and
  extraction of signals from the epoch of reionization using higher
  order one-point statistics}

\author[G. J. A. Harker et al.]{Geraint J. A. Harker,$^{1}$\thanks{E-mail:
  harker@astro.rug.nl} Saleem Zaroubi,$^{1}$ Rajat M. Thomas,$^{1}$
  Vibor Jeli\'c,$^{1}$ \newauthor
  Panagiotis Labropoulos,$^{1}$ Garrelt Mellema,$^{2}$ Ilian
  T. Iliev,$^{3}$ Gianni Bernardi,$^{1}$ \newauthor
  Michiel A. Brentjens,$^{4}$ A. G. de Bruyn,$^{1,4}$ Benedetta
  Ciardi,$^{5}$  Leon V. E. Koopmans,$^{1}$ \newauthor
  V. N. Pandey,$^{1}$ Andreas H. Pawlik,$^{6}$ Joop
  Schaye$^{6}$ and Sarod Yatawatta$^{1}$\\
  $^{1}$Kapteyn Astronomical Institute, University of Groningen, PO
  Box 800, 9700AV Groningen, the Netherlands\\
  $^{2}$Department of Astronomy, Alba Nova University Centre, Stockholm
  University, SE-106 91 Stockholm, Sweden\\
  $^{3}$Institute for Theoretical Physics, Winterthurerstrasse 190,
  CH-8057 Z\"urich, Switzerland\\
  $^{4}$ASTRON, Postbus 2, 7990AA Dwingeloo, the Netherlands\\
  $^{5}$Max-Planck Institute for Astrophysics,
  Karl-Schwarzschild-Stra\ss e 1, 85748 Garching, Germany\\
  $^{6}$Leiden Observatory, Leiden University, PO Box 9513, 2300RA
  Leiden, the Netherlands}
\begin{document}

\date{\today}

\maketitle

\begin{abstract}
Detecting redshifted 21cm emission from neutral hydrogen in the early
Universe promises to give direct constraints on the epoch of
reionization (EoR). It will, though, be very challenging to extract
the cosmological signal (CS) from foregrounds and noise which are
orders of magnitude larger. Fortunately, the signal has some
characteristics which differentiate it from the foregrounds and noise,
and we suggest that using the correct statistics may tease out
signatures of reionization. We generate mock datacubes simulating the
output of the Low Frequency Array (LOFAR) EoR experiment. These cubes
combine realistic models for Galactic and extragalactic foregrounds
and the noise with three different simulations of the CS. We fit out
the foregrounds, which are smooth in the frequency direction, to
produce residual images in each frequency band. We denoise these
images and study the skewness of the one-point distribution in the
images as a function of frequency. We find that, under sufficiently
optimistic assumptions, we can recover the main features of the
redshift evolution of the skewness in the 21cm signal. We argue that
some of these features -- such as a dip at the onset of reionization,
followed by a rise towards its later stages -- may be generic, and
give us a promising route to a statistical detection of reionization.
\end{abstract}

\begin{keywords}
cosmology: theory -- diffuse radiation -- methods: statistical -- radio lines: general
\end{keywords}

\section{Introduction}\label{sec:intro}

Between redshift $z\approx 20$ and $z\approx 6$ the Universe underwent
a transition from being almost entirely neutral to almost entirely
ionized (\citealt{BEN06}; \citealt*{FOB06}). This period, the epoch of
reionization (EoR), saw the first collapsed objects emit radiation
which heated and ionized the surrounding diffuse gas (additional
sources of heating and ionization, such as dark matter decay, have
also been considered; see, e.g., \citealt{VAL07}). Studying the
emission from this gas, in particular the redshifted 21cm line of
neutral hydrogen \citep*{FIE58,FIE59,HOG79,SCO90,KUM95,MAD97}, may therefore
tell us about the physics of these objects, and about structure
formation in a redshift range that has previously been explored rather
less directly. For example, quasar absorption spectra can constrain
the properties of the intergalactic medium towards the end of
reoinization \citep{FAN06}, while measurements of temperature and
polarization anisotropies in the cosmic microwave background provide
an integral constraint on the density of free electrons between the
observer and the surface of last scattering \citep[e.g.][]{DUN08}.

Several current and upcoming facilities (e.g.\ GMRT,\footnote{Giant
Metrewave Radio Telescope, http://www.gmrt.ncra.tifr.res.in/}
MWA,\footnote{Murchison Widefield Array,
http://www.haystack.mit.edu/ast/arrays/mwa/} LOFAR,\footnote{Low
Frequency Array, http://www.lofar.org/} 21CMA,\footnote{21 Centimeter
Arrat, http://web.phys.cmu.edu/\~{}past/} PAPER,\footnote{Precision
Array to Probe EoR, http://astro.berkeley.edu/\~{}dbacker/eor/}
SKA\footnote{Square Kilometre Array, http://www.skatelescope.org/})
will be sensitive to emission of the right wavelength to detect a
signal from neutral hydrogen during the EoR.  Significant
observational challenges must be overcome, however, before a
convincing detection can be made. The Galactic and extragalactic
foregrounds have a mean amplitude around \hbox{4--5} orders of
magnitude larger than the expected EoR signal (though their
fluctuations, which are the relevant quantity for an interferometer,
are only around three orders of magnitude larger; see, e.g.,
\citealt{SHA99}). Even closer to home, the signal is corrupted by the
ionosphere, radio frequency interference and instrumental
effects. Assuming all these factors can be dealt with, for realistic
integration times with the imminent generation of facilities the
random noise on the measurement per resolution element will still be a
few times larger than the signal.

The prospect of new observations of a poorly constrained period in the
Universe's history also poses a challenge to theorists: to model and
characterize the 21cm emission in such a way that it can be
meaningfully compared to the data
(e.g. \citealt*{BAR01,LOE01,CFW03,CSW03,BRO04};
\citealt{GLE06,ILI06,ILI08,ZAR07,THO08}). Given the observational
limitations listed above, it is unlikely that there will soon be clean
maps of the EoR signal with which to confront models. The first
detection of reionization from redshifted 21cm data will therefore be
of a statistical nature. This raises the question of precisely which
statistics to use: a question discussed by, e.g.,
\citet*{FUR04b,FUR04a,BHA05,GLE08}. In the first instance, they should
provide a clear indication of the global transition from a Universe
that is mostly neutral to one that is mostly ionized. Ideally, they
should be able to discriminate between different models describing the
more detailed progress of reionization. Most importantly, though, they
should be robust to the contamination introduced by the observing
process, and to the presence of high levels of noise.

We propose using the skewness of the one-point distribution of the
brightness temperature to study reionization, though we also consider
the prospects of other, similar statistics: the unnormalized third
moment and the kurtosis of the distribution. As we shall see below,
general arguments suggest that the skewness should be a strongly
evolving function of redshift during the EoR, and these arguments are
supported by simulations. Using these simulations, we generate
datacubes which also incorporate realistic models for the foregrounds,
instrumental response and noise levels expected for the LOFAR EoR
experiment. We generate residual images at each observed frequency by
attempting to remove the foregrounds using a fitting algorithm, then
study the properties of these residual images as a function of
redshift. If these residual images are denoised appropriately, we find
that we can indeed track the progress of reionization using the skewness.

In Section~\ref{sec:genapp} we introduce the skewness and explain how
it may help. In Section~\ref{sec:data} we give a brief description of
our models for the cosmological signal (CS), instrumental response,
foregrounds and noise. Then, in Section~\ref{sec:res}, we describe our
method for extracting the signal from datacubes which combine all
these components, and present our results. We discuss some possible
problems with and extensions to our methods in Section~\ref{sec:disc},
and finally we offer some conclusions in Section~\ref{sec:conc}.

\section{General approach}\label{sec:genapp}

We assume, with reasonable observational support, that the foregrounds
are smooth as a function of frequency, and exploit this in order to
extract the CS (\citealt{SHA99}; \citealt{DIM02};
\citealt{OH03}; \citealt*{ZAL04}). A simple way to imagine doing this
is first to estimate the foregrounds, either by fitting a smooth
function to the intensity as a function of frequency along each line
of sight, or by applying a filtering procedure. Subtracting this
estimate from the total yields fitting residuals which are an estimate
of the CS plus random noise.

If the variance of the noise is known, then at each frequency (or
equivalently at each redshift) we can attribute excess variance in the
fitting residuals over and above this level as coming from the
CS. This may yield a statistical detection of
emission from the epoch of reionization. \citet{JEL08} demonstrate
this method using the same foregrounds and noise levels as are used in
this work.

There are some difficulties with both parts of the above procedure.
Firstly, we may over-fit or under-fit, leading (respectively) to
underestimating or overestimating the residuals. This problem is
exacerbated near the ends of the frequency range. Secondly, the noise
properties may not be sufficiently well characterized. The variance of
the signal is expected to be only a small fraction of the variance of
the noise, and hence the latter must be very well known.

We therefore seek a statistic on the fitting residuals which will
detect the onset of reionization more robustly with respect to errors
in our estimates of the variance. A possible candidate is the
skewness, $\gamma_1$, defined in general for a continuous distribution
with probability density function $f(x)$ as
\begin{equation}
\gamma_1\equiv\frac{\mu_3}{\sigma^3}\equiv
\frac{\int_{-\infty}^{\infty}(x-\mu)^3f(x)\mathrm{d}x}
{\left(\int_{-\infty}^{\infty}(x-\mu)^2f(x)\mathrm{d}x\right)^{\frac{3}{2}}}
\label{eqn:skewdef}
\end{equation}
where $\sigma^2$ is the variance of the distribution and $\mu_3$ is
the third moment about its mean, $\mu$. A distribution which is mainly
concentrated at low $x$, but with a tail towards high $x$, will have
positive skewness. Similarly, a distribution with a tail extending to
low $x$ will have negative skewness, and a poor estimate of $\sigma$
cannot change its sign.

Rather than dealing with a continuous distribution, we will be
computing the skewness for images with $N$ pixels, in which the $i$th
pixel has a temperature $T_i$. Then the skewness may be expressed as
\begin{equation}
\gamma_1=\frac{\frac{1}{N}\sum_i\left(T_i-\bar{T}\right)^3}{\left(\frac{1}{N}\sum_i(T_i-\bar{T})^2\right)^{\frac{3}{2}}}
\label{eqn:skewdisc}
\end{equation}
where $\bar{T}$ is the mean temperature in that image and the sums are
over all pixels. In the case of residual images, if the foreground
fitting is unbiased, and if the noise is not skewed, then any
significant skewness remaining must come from the cosmological
signal. Moreover, we may expect some skewness in the cosmological
signal, which becomes very non-Gaussian once ionized bubbles appear in
large numbers during reionization.

To be concrete, making some reasonable assumptions and approximations
(that the optical depth is much less than unity, that the spin
temperature of the neutral hydrogen is much greater than the CMB
temperature, and that at these redshifts the Hubble parameter
$H(z)\approx\Omega_\mathrm{m}^{1/2}H_0(1+z)^{3/2}$), the difference,
$\delta T_\mathrm{b}$, between the brightness temperature of the 21cm
emission and the CMB is given by (\citealt{MAD97}; \citealt*{CIA03})
\begin{equation}
\frac{\delta
  T_\mathrm{b}}{\mathrm{mK}}=39h(1+\delta)x_\mathrm{HI}\left(\frac{\Omega_\mathrm{b}}{0.042}\right)\left[\left(\frac{0.24}{\Omega_\mathrm{m}}\right)\left(\frac{1+z}{10}\right)\right]^\frac{1}{2}
\label{eqn:dtb}
\end{equation}
where $\delta$ is the matter density contrast, $x_\mathrm{HI}$ is the
neutral hydrogen fraction, and the current value of the Hubble
parameter, \hbox{$H_0=100h\ \mathrm{km}\ \mathrm{s}^{-1}\
\mathrm{Mpc}^{-1}$}.

\begin{figure}
  \begin{center}
    \leavevmode \psfig{file=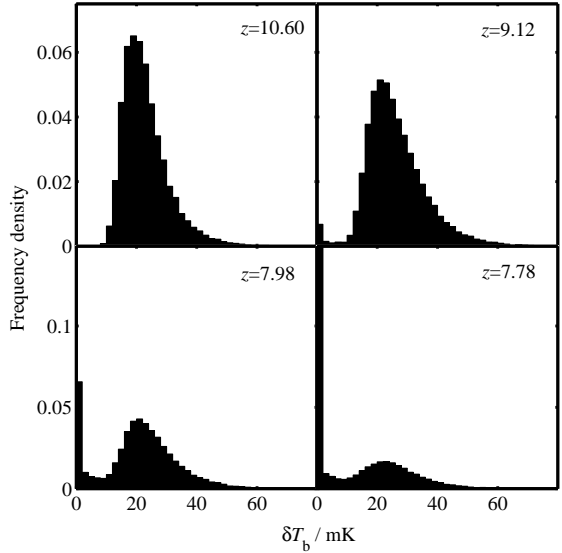,width=8cm}
    \caption{The distribution of $\delta T_\mathrm{b}$ in the f250C
    simulation (see Section~\ref{subsec:sigsim}) at four different
    redshifts, showing how the distribution evolves as reionization
    proceeds. Note that the y-axis scale in the top two panels is
    different from that in the bottom two panels. The delta-function
    at $\delta T_\mathrm{b}=0$ grows throughout this period while the
    rest of the distribution retains a similar shape. The bar for the
    first bin in the bottom-right panel has been cut off: approximately 58
    per cent of points are in the first bin at $z=7.78$.}\label{fig:onepoint}
  \end{center}
\end{figure}

At high redshift, when $x_\mathrm{HI}$ is close to unity everywhere,
the distribution of intensities is governed by the density field,
$1+\delta$. Initially this is nearly Gaussian, but develops a positive
skewness due to gravitational instabilities: see, for example,
\citet{PEE80}. This period is illustrated in the top left panel of
Fig.~\ref{fig:onepoint}, which shows the one-point distribution of
$\delta T_\mathrm{b}$ in one of our simulations (the f250C simulation;
see Section~\ref{subsec:sigsim}) at $z=10.6$, corresponding to an
observed frequency of $122.5\ \mathrm{MHz}$.
If reionization then takes place in patches, with large volumes
remaining mostly neutral while almost fully ionized bubbles form
around sources of ionizing photons, this has the effect of setting
$x_\mathrm{HI}=0$ (and so $\delta T_\mathrm{b}=0$) within the bubbles,
affecting the distribution of $\delta T_\mathrm{b}$ outside the
bubbles only weakly. So, in an idealized case, reionization takes
points from the distribution of $\delta T_\mathrm{b}$ and moves them
to a Dirac delta-function at zero. This has the effect of making the
skewness less positive; it may even become negative. The distribution
of $\delta T_\mathrm{b}$ at an early stage in this process ($z=9.12$,
or $140.3\ \mathrm{MHz}$) is shown in the top right panel of
Fig.~\ref{fig:onepoint}. By $z=7.98$ ($158.2\ \mathrm{MHz}$;
bottom-left panel) the two parts of the distribution are very
distinct. This may help to make it clear how the skewness can vanish:
the mean $\delta T_\mathrm{b}$ lies between these two peaks, and the
negative contribution to the skewness from points in the delta
function at zero may cancel with the positive contribution from points
to the right. At a later stage of reionization, when most of the
pixels in a noiseless map of $\delta T_\mathrm{b}$ at a given
frequency have values near zero, the points outside ionized bubbles
form a high-$\delta T_\mathrm{b}$ tail, giving the overall
distribution a strong positive skew. This can be seen in the bottom
right panel of Fig.~\ref{fig:onepoint} ($z=7.78$ or $161.8\
\mathrm{MHz}$). Note the short time between the third and fourth
panels: the later stages of reionization can progress rather quickly
as the number of ionizing sources can rise very rapidly, especially if
a major part is played by massive sources residing in haloes in the
exponential tail of the mass function \citep{JEN01}.

In this idealized case, then, the skewness as a function of redshift
should show a dip in the early stages of reionization, before growing
large in the later stages. Our aim in the subsequent sections of this
paper is to test if such a characteristic feature is indeed seen in
realistic simulations of reionization, and whether or not it can
provide a robust detection; or in other words, whether the effects of
foreground subtraction, noise, and instrumental corruption can mask or
mimic the signal.

\section{Simulations}\label{sec:data}

\subsection{Cosmological signal}\label{subsec:sigsim}

We use three simulations to estimate the CS. The
first and most detailed is the simulation labelled f250C by
\citet{ILI08}. The methodology behind this simulation is more fully
described by \citet{ILI06} and \citet{MEL06b}. The cosmological
particle-mesh code \textsc{pmfast} \citep*{MER05} was used to follow
the distribution of dark matter, using $1624^3$ particles on a
$3248^3$ mesh. The ionization fraction was then calculated in
post-processing using the radiative transfer and non-equilibrium
chemistry code \textsc{c$^2$-ray} \citep{MEL06a}. This takes place on
a coarser, $203^3$ mesh, and this is therefore the size used in this
work. The simulation box has a comoving size of $100\ h^{-1}\
\mathrm{Mpc}$. The cosmological parameters are close to those inferred
from the three-year {\it Wilkinson Microwave Anisotropy Probe} data
\citep[{\it WMAP}3: ][]{SPE07}, namely \hbox{($\Omega_\mathrm{m}$,
$\Omega_\Lambda$, $\Omega_\mathrm{b}$, $h$, $\sigma_8$,
$n$)}$=$\hbox{(0.24, 0.76, 0.042, 0.73, 0.74, 0.95)}.

A slightly different approach, detailed by \citet{THO09}, is used to
generate our other simulations. The dark matter distribution is
calculated using the \textsc{tree-pm} $N$-body code \textsc{gadget2}
\citep*{SPR01b,SPR05a}. Ionization is then calculated using a
one-dimensional radiative transfer code \citep{THO08}. The speed of
this approach means it is possible to study many more alternative
scenarios for the reionization process, while retaining good agreement
with more accurate, three-dimensional radiative transfer
simulations. We will show results from two different simulations. In
both cases, the dark matter simulation uses $512^3$ dark matter
particles in a box of comoving size $100\ h^{-1}\ \mathrm{Mpc}$, with
\hbox{($\Omega_\mathrm{m}$, $\Omega_\Lambda$, $\Omega_\mathrm{b}$,
$h$, $\sigma_8$, $n$)}$=$\hbox{(0.238, 0.762, 0.0418, 0.73, 0.74,
0.951)}. While the simulations contain no baryons, this value of
$\Omega_\mathrm{b}$ was used to generate the initial power
spectrum. These parameters give them lower resolution than the dark
matter part of the f250C simulation, meaning that low mass sources are
not resolved and are neglected. In one of these simulations we assume
that the Universe is reionized by QSOs, and in the other by
stars. These two simulations are labelled `T-QSO' and `T-star'
respectively. The former should not be affected too seriously by the
lack of resolution, since QSOs do not reside in low-mass haloes. In
the latter, the geometry of reionization may be altered: compared to a
higher resolution simulation, larger ionized bubbles may form at a
given global star formation rate, for example. As we shall see below,
`T-star' shows rather different characteristics from the f250C
simulation, despite the fact that stars provide the ionizing photons
in both cases. This illustrates the uncertainties involved in
modelling the physics of reionization, in selecting the source
populations and finding their distribution in space, and in choosing
the approximations required to make the calculations tractable. We do
not analyze the differences between the simulations in great detail
here; rather, we use the different simulations to provide a variety of
plausible scenarios with which to test our signal extraction
techniques.

The above calculations all take place in periodic boxes. The final
step in generating a datacube is to take a series of simulation
snapshots at different redshifts, and interpolate between them to
produce a spatial slice at each observed frequency. This procedure,
which is described in detail by \citet{THO09} and \citet{MEL06b}, is
analogous to the generation of lightcone output to compare to galaxy
surveys \citep{EVR02}. In performing this conversion from a position
in a periodic box to a redshift, we take account of the peculiar
velocities; that is, our datacubes are in redshift space, as will be
the case for the observational data. As emphasized by \citet{MEL06b},
who were the first to include peculiar velocity distortions in a
redshifted 21cm context based on detailed simulations, these effects
can be important. At linear scales, the redshift space distortions
have the effect of enhancing density fluctuations along the line of
sight \citep{KAI87,BHA05}.

\subsection{Foregrounds, noise and instrumental effects}\label{subsec:fg}

We use the foreground simulations described in detail by
\citet{JEL08}. These incorporate contributions from Galactic diffuse
synchrotron and free-free emission, and supernova remnants. They also
include extragalactic foregrounds from radio galaxies and radio
clusters. The foreground maps cover an area $5^{\circ}\times
5^{\circ}$ on the sky, which corresponds to the area of one LOFAR EoR
window. We also adopt the frequency-dependent noise levels given by
\citet{JEL08}. The noise is described in more detail below, when we
introduce each of our two noise models.

In Section~\ref{sec:res}, we will give results obtained by combining
the CS, foregrounds and noise, and then attempting to
extract the CS. We also attempt the more difficult,
and more realistic task of extracting the signal given such a datacube
corrupted by the instrumental response (so-called `dirty maps'). These
instrumental effects will be described in more detail by
\citet{LAB09}. In brief, at each observed frequency, each pair of
LOFAR stations gives an estimate of the Fourier transform of the sky
brightness along a track of points (a \uv track) in Fourier space (the
\uv plane). Translated into configuration space, this means that our
images of the sky are convolved with a complicated point spread
function (the `dirty beam'). Moreover, because the origin of the
Fourier plane is not sampled, interferometer measurements are not
sensitive to the mean brightness. A more detailed introduction to such
datacubes, and their relation to the power spectrum of emission from
the EoR, is given by \citet{MOR04}.

\section{Results}\label{sec:res}

\subsection{Skewness in perfect data}

\begin{figure}
  \begin{center}
    \leavevmode \psfig{file=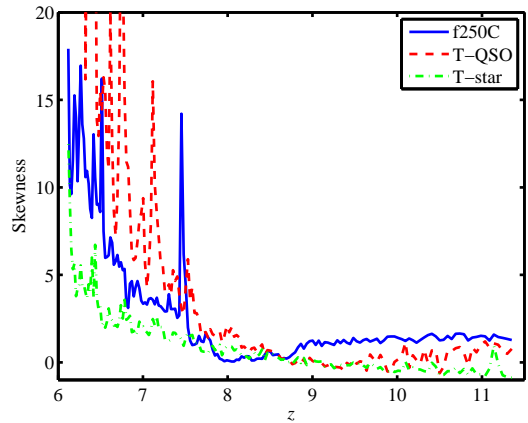,width=8cm}
    \caption{The evolution of the skewness of the distribution of
    $\delta T_\mathrm{b}$ as a function of redshift in our three
    simulations. Each point corresponds to the skewness of the
    one-point distribution in a slice through a datacube at the given
    redshift. The most striking feature here is the steep rise in
    skewness at low redshift. The details at high redshift can be seen
    more clearly in Fig.~\ref{fig:skewandmean}.}\label{fig:skewo}
  \end{center}
\end{figure}

The evolution of skewness in our three simulations, uncorrupted by
noise or foregrounds, for the redshift range observable by LOFAR, is
given in Fig.~\ref{fig:skewo}.
The most obvious feature here, common to all three simulations, is the
rise in the skewness at low redshift, due to the high-$\delta
T_\mathrm{b}$ tail of points with some remaining neutral hydrogen. At
higher redshift there are some differences, however. In f250C there is
a clear dip at $z\sim7.8$--$9$, whereas this is not so obvious in either of
the other simulations. For the T-QSO simulation, though, and more
noticeably for the T-star simulation, the skewness is negative in certain
redshift slices. All the simulations show large spikes in skewness at
low redshift. At these redshifts, a slice may have only one or a
few regions from which there is significant emission. This leads to
large variation between slices. A slice in which one small region
produces a high-emission tail in the one-point distribution, and which
is surrounded by more uniform slices, shows up as a spike in the
evolution of the skewness.

\begin{figure}
  \begin{center}
    \leavevmode \psfig{file=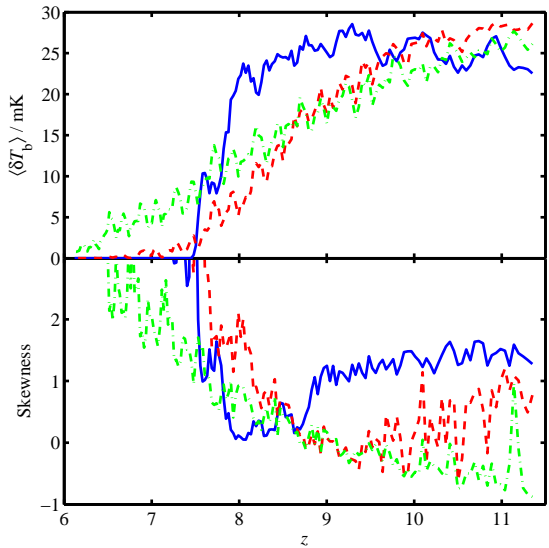,width=8cm}
    \caption{The evolution of the mean differential brightness
    temperature (top panel) and of the skewness (bottom panel) in the
    three simulations. The colours and styles of the lines are the
    same as in Fig.~\ref{fig:skewo}. The top panel shows that
    reionization is extended to very different degrees in the
    different simulations. We cut off the large values of the skewness
    in the bottom panel so that the dip in skewness at high redshift
    can be seen more clearly than in
    Fig.~\ref{fig:skewo}.}\label{fig:skewandmean}
  \end{center}
\end{figure}

The lower-skewness region of the plot is shown in more detail in
Fig.~\ref{fig:skewandmean}, where we also compare the evolution of the
skewness to that of the mean.
Whilst the mean value of $\delta T_\mathrm{b}$ is unobservable with an
interferometer such as LOFAR, this serves to illustrate the progress
of reionization. Clearly the process is much more rapid in f250C than
in the T-QSO simulation, which in turn is more rapid than the T-star
simulation. In Fig.~\ref{fig:skewandmean} a dip in the skewness is
discernible in T-QSO, but is much less obvious than the dip in f250C
because of fluctuations at high redshift, and because it spans a wider
range in redshift due to the more extended reionization. This may be
an indication that a more extended reionization process will be harder
to detect using the skewness. The T-star simulation provides some
hope, however. Despite an even more gradual reduction in the mean
differential brightness temperature, the skewness is negative for
quite a large range in redshift. Because the density field is
positively skewed, a negative skewness is a clear signature of
reionization.  It is therefore possible that the skewness could
provide a detection even in the case of very extended reionization.

\subsection{Extracted skewness}\label{subsec:extracskew}

We now proceed to test the possibilities for signal extraction using
the skewness, starting with a rather more optimistic case than will be
encountered with the actual LOFAR EoR experiment. We first note that
the $5^{\circ}\times 5^{\circ}$ field of one LOFAR EoR window
corresponds to a distance of approximately $800\ \mathrm{Mpc}\times
800\ \mathrm{Mpc}$ (comoving) at $z=10$ in our assumed cosmology. At
each redshift we therefore tile copies of the simulation to produce a
slice of the correct size, then interpolate this onto a $256^2$
mesh. Since each pixel will be affected differently by foregrounds and
noise, and since we consider only one-point statistics, we do not
anticipate that this will strongly affect our conclusions. To produce
our datacube, we add the simulated foregrounds described in
Section~\ref{subsec:fg} to the CS, then smooth each
slice using a Gaussian kernel to the estimated LOFAR resolution of
$\approx 4\ \mathrm{arcmin}$. We then add uncorrelated random noise as
described by \citet{JEL08}. The noise has an {\it rms} of $52\
\mathrm{mK}$ at $150\ \mathrm{MHz}$ and has two contributions: a
frequency dependent component coming from the sky, which scales as
$\nu^{-2.55}$, and a frequency-independent part from the
receivers. The noise on each image pixel is independent. In reality,
this will not be the case: rather, the noise on individual
visibilities will be independent. We will tackle this more difficult
case with realistic noise and a non-Gaussian point spread function
below. Spatial slices are separated by $0.5\ \mathrm{MHz}$ in
frequency, $\nu$. At $150\ \mathrm{MHz}$ this corresponds to a
difference in redshift, $\Delta z\approx 0.03$, the slices having a
comoving thickness of approximately $7\ h^{-1}\ \mathrm{Mpc}$.

Once we have a datacube with EoR signal, foregrounds and noise, we fit
a third-order polynomial in $\log\nu$ to each pixel. We have
experimented with using different functional forms, but find that so
long as we obtain a visually reasonable fit, our results for the
skewness do not change enough to affect our conclusions. Measurements
of the variance are rather more sensitive to under- and over-fitting,
which demonstrates the importance of understanding the foregrounds
well, and of using robust statistics. It is also possible to estimate
the foregrounds by removing noise using a filtering procedure. While
this requires fewer assumptions about the nature of the foregrounds,
it tends to over-fit.

\begin{figure}
  \begin{center}
    \leavevmode \psfig{file=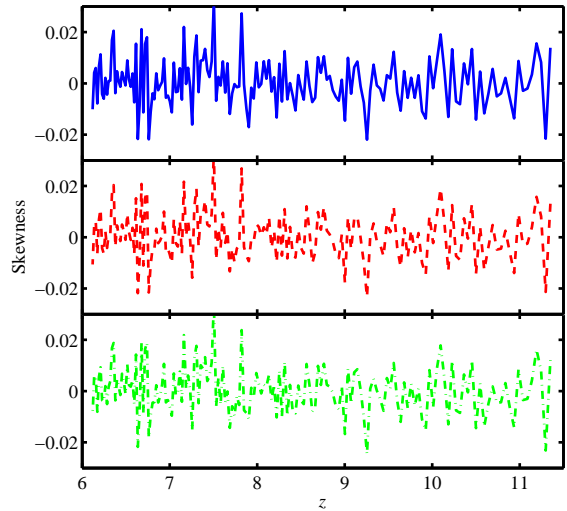,width=8cm}
    \caption{Skewness of the fitting residuals from datacubes with
    uncorrelated noise. The same noise is used in each cube. The
    colours and styles of the lines are the same as in
    Fig.~\ref{fig:skewo}: in the top panel, the component of the
    datacube from the CS comes from f250C; in the
    middle panel, from the T-QSO simulation; and in the bottom
    panel, from the T-star simulation. The similarity between the
    three panels arises because the noise realization (and the
    foregrounds) is the same for each panel, and dominates over the
    CS.}\label{fig:rskew3}
  \end{center}
\end{figure}

After a fit has been obtained, it is subtracted from the total,
leaving residuals which are an estimate of the CS
plus the noise.  The skewness of these residuals as a function of
redshift is shown in Fig.~\ref{fig:rskew3}.
While the CS which goes into the datacube is
different for each panel of the plot, the noise and foregrounds are
the same. This accounts for the fact that the skewness as a function
of redshift shows very similar features in each panel.  At first sight
this seems rather discouraging, with the desired signal totally
dominated by fitting errors (FEs) and noise.

\begin{figure}
  \begin{center}
    \leavevmode \psfig{file=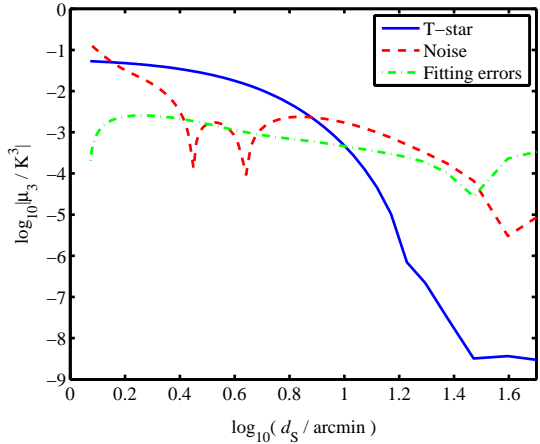,width=8cm}
    \caption{The effect of smoothing scale on the absolute value of
    the third moment of the one-point distribution, $|\mu_3|$ (defined
    in Equation~\ref{eqn:skewdef}), for the three different components
    of the residual maps at $115\ \mathrm{MHz}$ ($z=11.35$). The
    width, $d_\mathrm{S}$, of the Gaussian distribution which forms
    our smoothing kernel is defined here as the distance between
    $x=\pm\sigma$ for a distribution with standard deviation
    $\sigma$. The CS (blue, solid line) is from the
    T-star simulation, in which the skewness of the one-point
    distribution is negative at this frequency. The red, dashed line
    corresponds to the noise (where here the noise on each pixel in
    the unsmoothed map is independent), and the green, dot-dashed line
    to the fitting errors, by which we mean the difference between the
    foregrounds cube and the polynomial fit to the full datacube. The
    value for the foregrounds themselves, before fitting, is several
    orders of magnitude larger.}\label{fig:scaledep}
  \end{center}
\end{figure}

The situation can be improved, however. At each frequency, we may
denoise the residual image by smoothing. This is possible because our
images are oversampled, with more than one pixel per resolution
element, and because there are no pixel-to-pixel correlations in the
noise (by construction). While this is clearly unrealistic, it serves
as a prototype for the more difficult denoising step when we consider
the dirty maps. The effect of smoothing on the different components of
the residual maps -- these components being the CS, FEs and noise --
is illustrated in Fig.~\ref{fig:scaledep}, in which we show how the
absolute value of the third moment of the one-point distribution of
these components, $|\mu_3|$, is affected when they are smoothed with
windows of different size.
We show the result for the slice of the T-star datacube at $115\
\mathrm{MHz}$, corresponding to a redshift of $11.35$. This slice is
chosen because the skewness of the T-star simulation is significantly
negative here; we get similar results with the other simulations if we
choose an appropriate slice in which the skewness is significantly
different from zero. When the smoothing window is very narrow, so that
there is almost no smoothing, $|\mu_3|$ for the noise exceeds the
value for the CS. This occurs even though
$\langle\mu_3^\mathrm{noise}\rangle=0$ (where the expectation is taken
over different noise realizations), simply because the noise {\it rms}
is so much larger than that of the (significantly skewed) cosmological
signal. As the size of the smoothing window is increased,
$|\mu_3^\mathrm{noise}|$ drops much more quickly than
$|\mu_3^\mathrm{CS}|$ since the smoothing averages together
uncorrelated noise pixels, but correlated signal pixels.  At large
scales, the signal also becomes uncorrelated, so its small {\it rms}
means that $|\mu_3^\mathrm{CS}|<|\mu_3^\mathrm{noise}|$ once more. The
scale at which $|\mu_3^\mathrm{CS}|$ exceeds $|\mu_3^\mathrm{noise}|$
by the greatest amount in this residual map is approximately $3$--$4$
$\mathrm{arcmin}$. In the case of the fitting errors,
$|\mu_3^\mathrm{FE}|$ shows less variation as the smoothing scale is
changed than either of the other components. If the foreground fitting
were completely unbiased, we might expect that any errors in the
fitting would be Gaussian and caused entirely by noise, and so this
component of the residual images would behave similarly to the noise
component. The fact that the skewness in the fitting errors appears to
come partly from large scales suggests that bias in the fitting may
allow some leakage through from the foregrounds themselves, which are
correlated on large scales. If the skewness of the foregrounds is
larger than we have assumed, therefore, we will need to fit them more
accurately in addition to exploiting this scale dependence. In the
present case, $|\mu_3^\mathrm{FE}|$ is similar to
$|\mu_3^\mathrm{noise}|$ at the scale at which the latter is dwarfed
by the contribution from the CS.

\begin{figure}
  \begin{center}
    \leavevmode \psfig{file=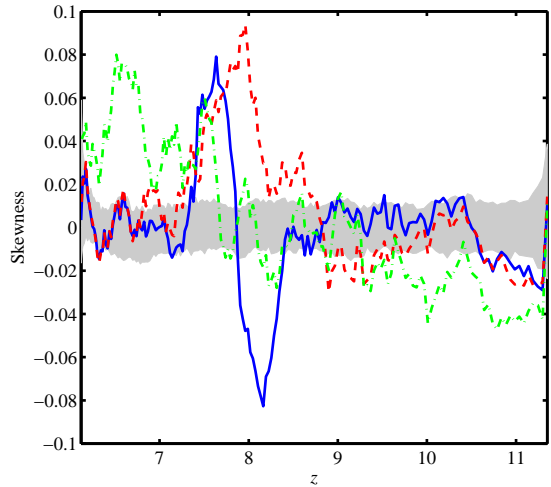,width=8cm}
    \caption{Skewness of the fitting residuals from datacubes with
    uncorrelated noise, but in which the residual image has been
    denoised by smoothing at each frequency before calculating the
    skewness. The three lines correspond to the three simulations,
    with colours and line styles as indicated in
    Fig.~\ref{fig:skewo}. Each line has been smoothed with a moving
    average (boxcar) filter with a span of nine points. The grey,
    shaded area shows the errors, estimated using 100 realizations of
    the noise. The fact that the lines differ so much is in marked
    contrast to Fig.~\ref{fig:rskew3}, and shows the impact of
    smoothing the residual images to suppress the
    noise.}\label{fig:rskew3s}
  \end{center}
\end{figure}

In practice, to extract the skewness as a function of frequency we
make the natural choice of smoothing scale, using the same kernel as
was used to degrade the signal and foregrounds to the resolution of
the telescope. We then compute the skewness in each slice as
before. The skewness as a function of frequency for each datacube,
after following this procedure, is shown in Fig.~\ref{fig:rskew3s}.
To improve the clarity of the plot, each line is smoothed by taking a
moving average with a span of nine points (a boxcar filter). To
estimate the error, we generate 100 datacubes containing the
foregrounds and with different realizations of the noise, but with no
CS present. We feed each cube through our fitting and
smoothing procedure, calculate the skewness as a function of redshift,
and smooth this function with a moving average filter just as for the
cubes containing a signal. The range between the 16th and 84th
percentile of the skewness for these realizations is shown as the
light grey shaded area in the figure.

One can see from Fig.~\ref{fig:rskew3s} that this smoothing procedure
allows us to extract a significant signal, despite making only rather
general assumptions about the scale at which features due to the
signal, instrument and noise are important. The result for f250C
(blue, solid line) is most striking, with rapid transitions in the
skewness in the range $z\approx 7.5$--$8.5$. The position of the dip
corresponds to the position of the dip in the uncorrupted simulation
shown in Fig.~\ref{fig:skewo}. While the skewness continues to rise in
the original simulation, however, for the extracted signal it returns
to zero at low redshift. This is because the variance of the
CS becomes very small at low redshift. In the
uncorrupted simulation, this allows the skewness to grow very
large. In the residual images, however, the variance of the noise and
fitting residuals comes to dominate, even after smoothing, which
drives the skewness towards zero. We return to this point below when
we consider alternative statistics. The extracted signal for the other
two simulations shows the behaviour one might expect: the T-QSO
simulation (red, dashed line) shows only a weak dip in skewness, but a
strong peak due to the rapid rise in skewness for the uncorrupted
simulation at $z\lesssim 8.5$. The T-star simulation, meanwhile, shows
a gradual rise in skewness throughout the redshift range, with
significant non-zero skewness detected for $z\gtrsim 9.5$ and
$z\lesssim 7.5$.

\subsection{Skewness from dirty maps}

We now move on to an analysis of the so-called `dirty maps'. To
generate these, we first add together the unsmoothed foreground and
signal cubes, where the latter have been tiled, as before, to produce
maps of the right angular size. Each slice is then corrupted by the
instrumental response. We achieve this in practice by Fourier
transforming the image, multiplying by the sampling function
(calculated on a grid with the same number of points as the image),
and then applying the inverse Fourier transform. This is equivalent to
convolving each image with the point spread function (PSF) of the
instrument. We make the simplifying assumption that the sampling
function does not change with frequency. If the \uv plane is uniformly
filled, this should not be excessively optimistic. It could be
enforced in practice by discarding high-$k$ data so that equivalent
(and completely filled) areas of the \uv plane are retained in each
frequency band. The noise is dealt with slightly differently. We
consider pixels in the \uv plane where the sampling function is
non-zero to be encompassed by our \uv coverage, and we generate
uncorrelated Gaussian noise at each such pixel. Pixels outside our \uv
coverage are set to zero. We (inverse) Fourier transform to return to
the image plane, then normalize this `noise image' such that it has
the correct {\it rms}. This procedure yields noise that is almost
uncorrelated between independent resolution elements (though the noise
on adjacent pixels is correlated).

We fit out the foregrounds in the dirty cubes in the same way as
before. The skewness of the residual images exhibits the same problem
seen in Fig.~\ref{fig:rskew3}, being dominated by the noise. In this
case, the smoothing procedure used above would not be expected to
help, since the noise is correlated on the scale of our smoothing
kernel. In addition, since our resolution is comparable to the scale
of features in the original signal, using a broader kernel simply
washes out the signal as well as the noise. We therefore require a
more sophisticated denoising scheme.

\subsubsection{Wiener filtering}

With the results of Section~\ref{subsec:extracskew} in mind, we use
the differing correlation properties of the signal and noise in our
extraction. To be explicit, suppose that we write the residuals as a
vector $\boldsymbol{d}$, where $d_i$ is the residual at the $i$th pixel of
a map at a given frequency. We relate $\boldsymbol{d}$ to the image from
the uncorrupted simulation, $\boldsymbol{s}$, by
\begin{equation}
\boldsymbol{d}=\mathbfss{R}\boldsymbol{s}+\boldsymbol{\epsilon}\quad .
\label{eqn:deqrspe}
\end{equation}
The matrix $\mathbfss{R}$ encodes the convolution of the signal with the
PSF, while $\boldsymbol{\epsilon}$ represents the noise. We neglect any
contribution to $\boldsymbol{\epsilon}$ coming from errors in the fitting
procedure, so we can assume that the correlation matrix of the noise,
$\mathbfss{N}=\langle\boldsymbol{\epsilon\epsilon}^\dag\rangle$, is
known (here, $\boldsymbol{\epsilon}^\dag$ is the conjugate transpose of
$\boldsymbol{\epsilon}$).

We consider a very optimistic situation for extracting the skewness,
which occurs if the correlation matrix of the signal,
$\mathbfss{S}=\langle\boldsymbol{ss}^\dag\rangle$, is also known. We can
then perform a Wiener deconvolution on each residual image to recover
an estimate of the CS. That is, we compute
$\hat{\boldsymbol{s}}=\mathbfss{F}\boldsymbol{d}$ where the Wiener filter $\mathbfss{F}$ is
given by
\begin{equation}
\mathbfss{F}=\mathbfss{SR}^\dag(\mathbfss{RSR}^\dag +\mathbfss{N})^{-1}
\label{eqn:wf}
\end{equation}
\citep[see, e.g.,][]{ZAR95}. In the absence of noise, this procedure
reduces to an ideal inverse filter that estimates the orignal image
before corruption by the PSF. In the presence of noise, the Wiener
filter suppresses power in the image at those values of
$\boldsymbol{k}$ for which the signal-to-noise ratio (SNR) is low,
while retaining power for modes where the SNR is high (see, e.g.,
\citealt{NR86}). The algorithm is optimal in the least-squares
sense.

\begin{figure}
  \begin{center}
    \leavevmode \psfig{file=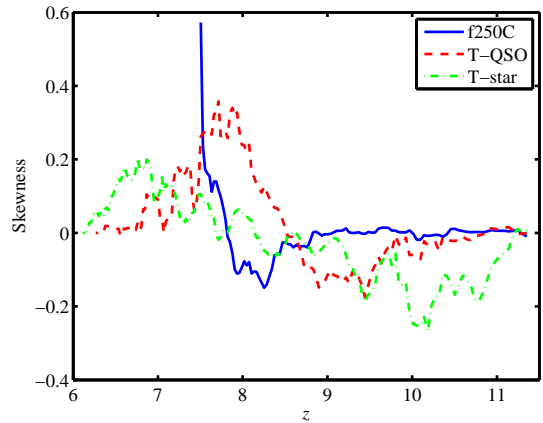,width=8cm}
    \caption{Skewness of deconvolved images as a function of
    redshift. The dirty maps are generated by convolving the
    CS and foregrounds with the instrumental
    response, and adding noise with a realistic correlation
    matrix. The CS is then estimated by fitting out
    the foregrounds and applying a Wiener deconvolution to the
    residuals, assuming that the PSF and the correlation matrices of
    signal and noise are known. Line colours and styles are the same
    as in Fig.~\ref{fig:skewo}. The deconvolution (as opposed to the
    simple smoothing which was sufficient for the uncorrelated noise
    used in Fig.~\ref{fig:rskew3s}) is unstable when the cosmological
    signal becomes very small, which prevents us from estimating the
    errors in the same way as for
    Fig.~\ref{fig:rskew3s}.}\label{fig:skewdb}
  \end{center}
\end{figure}

The skewness of these deconvolved images as a function of redshift is
shown in Fig.~\ref{fig:skewdb}.
Comparing to Figs.~\ref{fig:skewo} and~\ref{fig:skewandmean}, one can
see that this procedure gives excellent results, recovering the
general trends in skewness seen in the original simulations. Indeed,
using an optimal filter with precise knowledge of the signal and noise
properties means that we recover larger values for the skewness than
were seen after applying the simple smoothing to the uncorrelated
noise case of Section~\ref{subsec:extracskew}
(Fig.~\ref{fig:rskew3s}). We can realistically expect a situation
intermediate between the results of Figs.~\ref{fig:rskew3}
and~\ref{fig:skewdb}. The lines representing f250C and the T-QSO
simulation do not extend all the way to $z=6$ in
Fig.~\ref{fig:skewdb}. At these redshifts, the variance in the
CS is so small compared to that of the noise that the
deconvolution becomes unstable. For the same reason, we cannot
estimate the errors in the same way as for Fig.~\ref{fig:rskew3s}
(that is, by generating realizations with no CS at
all). None the less, the errors can be inferred to be small since we
do not see the same effect as in Fig.~\ref{fig:rskew3s}, in which the
extracted signal from the datacube generated with the f250C and T-QSO
simulations is very similar at high and low redshift, being dominated
by noise.

\begin{figure}
  \begin{center}
    \leavevmode \psfig{file=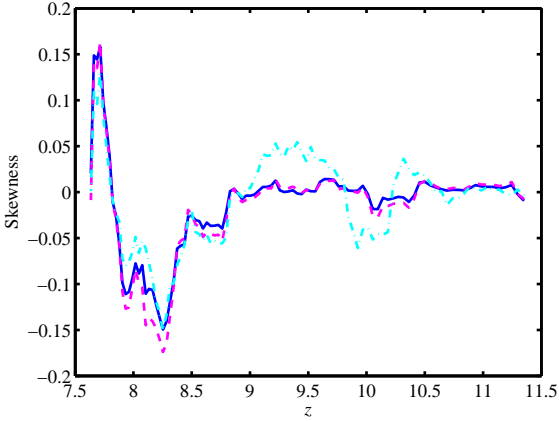,width=8cm}
    \caption{The skewness as a function of redshift recovered from the
    f250C simulation by Wiener deconvolution, using three different
    assumptions for the correlation matrix, $\mathbfss{S}$, of the
    original signal. For the solid, blue line we use the correct
    correlation matrix, $\mathbfss{S}_\mathrm{f250C}(z)$. For the
    dashed, magenta line we use
    $\frac{1}{2}\mathbfss{S}_\mathrm{f250C}(z)$, and for the dot-dashed,
    cyan line we use the correlation matrix from the T-star
    simulation.}\label{fig:diffcorr}
  \end{center}
\end{figure}

An obvious objection to the method presented here is that if the
correlation matrix of the CS is known, this means
that we have already detected a signal from the EoR, so higher-order
statistics are not required to extract it. The force of this objection
depends on how good an estimate of the correlation matrix of the
signal is required for the deconvolution to give an acceptable
result. We present a test of this in Fig.~\ref{fig:diffcorr}.
The three lines in the figure show the skewness extracted from the
f250C residuals using three different assumption for the correlation
matrix, $\mathbfss{S}$, used in the Wiener deconvolution, The solid,
blue line shows, for reference, the skewness extracted when we use the
correct correlation matrix $\mathbfss{S}_\mathrm{f250C}(z)$, calculated
from the original simulation, in performing the deconvolution. The
dashed, magenta line shows the extracted skewness when we use
$\frac{1}{2}\mathbfss{S}_\mathrm{f250C}(z)$ instead. Underestimating the
correlation matrix by a factor of two clearly has only a minimal
effect on the extraction of the skewness. Finally the dot-dashed, cyan
line shows the result when we use the correlation matrix of the T-star
simulation. Even though the redshift evolution of the two simulations
is very different, the dip and peak in the skewness at \hbox{$z\approx
7.5$--$8.5$} are recovered, though it is not clear that they can be
easily distinguished from the spurious variations at high
redshift. This preliminary result is encouraging, but it would be
preferable to use a correlation matrix estimated from the data
themselves. For example, writing Equation~\ref{eqn:wf} in terms of the
correlation matrix of the residuals,
$\mathbfss{D}=\langle\boldsymbol{dd}^\dag\rangle$, it becomes
\begin{equation}
\mathbfss{F}=\mathbfss{R}^{-1}(\mathbfss{D}-\mathbfss{N})\mathbfss{D}^{-1}\quad .
\label{eqn:wfd}
\end{equation}
Unfortunately this filter does not work so well in practice, since the
convolution means that important small-scale information present in
$\mathbfss{S}$ is not present in $\mathbfss{D}$. Since estimation of
the signal correlation matrix is intimately connected with techniques
for power spectrum estimation, we defer further investigation to a
future paper.

\subsubsection{Foreground subtraction effects}

\begin{figure}
  \begin{center}
    \leavevmode \psfig{file=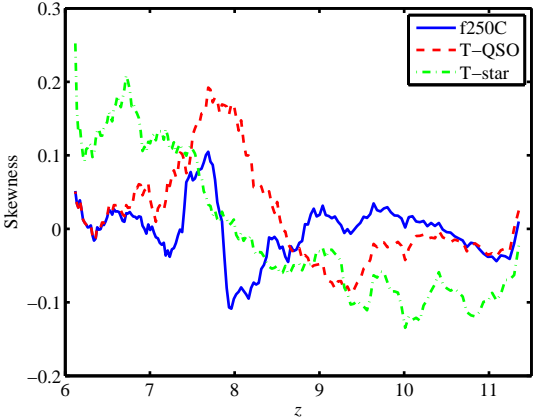,width=8cm}
    \caption{Skewness of dirty maps with realistic noise, but in which
    no foregrounds have been added. This is equivalent to a case in
    which we achieve perfect removal of the foregrounds.}\label{fig:skewnofg}
  \end{center}
\end{figure}

Part of the need for sophisticated denoising techniques comes from the
fact that imperfect fitting of the foregrounds introduces errors into
our residual images. We illustrate this in Fig.~\ref{fig:skewnofg}, in
which we show the skewness of the dirty maps to which we have added
realistic noise, but no foregrounds.
The main trends in the evolution of the skewness are clearly
visible. We have seen in Fig.~\ref{fig:rskew3} that even uncorrelated
noise prevents us from recovering a signal if we first have to
subtract the foregrounds. Therefore it is the combination of bright
foregrounds and structured noise which prevents us from extracting the
skewness without some assumptions about the properties of the
signal. It will therefore be important to develop further our
techniques for foreground subtraction, and to test how sensitive they
are to variations in our models for the foregrounds.

\subsection{Skewness from Fourier space data}

The observable quantity for an interferometer is the radio visibility:
a quantity that lives in Fourier space.  It also seems natural,
therefore, to try to extract a signal from the EoR without
transforming to the image plane first (though we note that some stages
of the analysis which must be completed to produce the clean datacube
from which we attempt to extract a signal, such as the subtraction of
bright point sources, are also carried out in the image plane). For
example, \citet{DAT07} have presented a formalism to search for
bubbles in 21cm data using a statistic on the visibilities.

Unfortunately, it is not convenient to calculate the skewness from
Fourier space data. The case for the moments of the density field is
well known: the variance of the overdensity field,
$\langle\delta^2\rangle$, is equal to the two-point correlation
function evaluated at zero separation, $\xi(0)$, which in turn is
equal to an integral over the power spectrum of fluctuations.
Similarly, the third moment $\langle\delta^3\rangle$ is equal to
$\zeta(0,0)$ where
$\zeta(\boldsymbol{r}_1,\boldsymbol{r}_2)\equiv
\langle\delta(\boldsymbol{x})\delta(\boldsymbol{x}+\boldsymbol{r}_1)
\delta(\boldsymbol{x}+\boldsymbol{r}_2)\rangle$.
Then we have
\begin{equation}
\zeta(0,0)=\int\mathrm{d}^3\boldsymbol{k}'\mathrm{d}^3\boldsymbol{k}''
B(\boldsymbol{k}',\boldsymbol{k}'',-\boldsymbol{k}'-\boldsymbol{k}'')
\end{equation}
where $B$ is the bispectrum, defined by
\begin{equation}
\langle\tilde{\delta}(\boldsymbol{k}_1)\tilde{\delta}(\boldsymbol{k}_2)
\tilde{\delta}(\boldsymbol{k}_3)\rangle\equiv
\delta_\mathrm{D}(\boldsymbol{k}_1+\boldsymbol{k}_2+\boldsymbol{k}_3)B(\boldsymbol{k}_1,\boldsymbol{k}_2,\boldsymbol{k}_3)
\end{equation}
where $\tilde{\delta}(\boldsymbol{k})$ is the Fourier counterpart of
$\delta(\boldsymbol{x})$ and $\delta_\mathrm{D}$ is the Dirac
delta-function. The bispectrum is a rather unwieldy object to grapple
with in this context. Foreground extraction is also problematic: while
the foregrounds remain smooth as a function of frequency at a given
\uv point, the angular scale sampled by that point is also a function of
frequency. Further consideration of Fourier space statistics is
therefore beyond the scope of this paper.

\subsection{Alternative statistics}

\begin{figure}
  \begin{center}
    \leavevmode \psfig{file=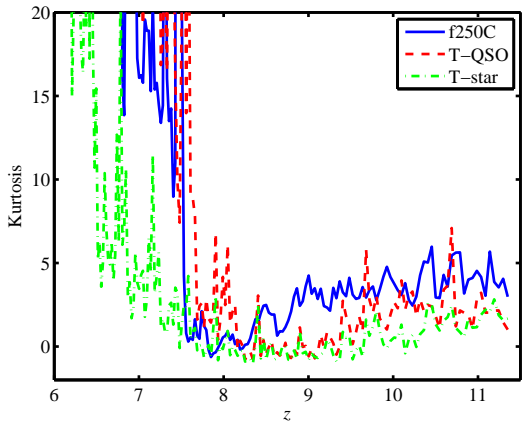,width=8cm}
    \caption{The evolution of the kurtosis of the distribution of
    $\delta T_\mathrm{b}$ as a function of redshift in our three
    simulations. Some of the qualitative features are similar to those
    seen in the skewness in Fig.~\ref{fig:skewo}, though for slightly
    different reasons which are explained in the main text. In the
    definition of kurtosis used here, a Gaussian distribution has a
    kurtosis of zero. As in the lower panel of
    Fig.~\ref{fig:skewandmean} we cut off the very large values at low
    redshift, in order to make the high redshift region
    clearer.}\label{fig:kurto}
  \end{center}
\end{figure}

The potential for using the skewness of the CS to
help in the extraction invites the question of whether other one-point
statistics, such as the kurtosis, could also be useful. We define the
kurtosis here as $\mu_4/\sigma^4-3$ where $\mu_4$ is the fourth
central moment of the distribution, and we subtract $3$ so that a
Gaussian distribution has a kurtosis of zero. In fact the kurtosis
does evolve strongly in the signal simulations. As was the case for
the skewness, it is not difficult to see why. While the brightness
temperature traces the density field it retains a kurtosis similar to
that of a Gaussian distribution. The formation of bubbles then
produces a bimodal distribution with no strong central peak, so the
kurtosis, a measure of the `peakiness' of the distribution, is
reduced. In the final stages of reionization, the distribution becomes
strongly peaked around zero, with a tail of points with strong
emission, leading to a large kurtosis. The evolution of the kurtosis
is therefore qualitatively similar to that of the skewness. We see all
these stages in our simulations, and the evolution of kurtosis as a
function of redshift is shown in Fig.~\ref{fig:kurto}.
Unfortunately these trends seem to be much harder to recover than in
the case of skewness. While we can weakly recover the dip for the case
of uncorrelated noise in the image plane, using a similar procedure to
that used for Fig.~\ref{fig:rskew3s}, we cannot recover the
low-redshift peak. For the dirty maps, the dip in kurtosis cannot be
seen even after the optimistic deconvolution used in
Fig.~\ref{fig:skewdb}.

Because most of the variance in the unsmoothed residual images comes
from the noise and fitting errors, which one might expect not to be
skewed, we have investigated whether or not the unnormalized third
moment, $\mu_3$, might perform better than the skewness,
$\mu_3/\sigma^3$, for signal extraction. In fact we find that they
perform similarly, but the results for the skewness tend to be easier
to interpret. This is because to calculate the errors we must still
estimate the expected spread in $\mu_3$ from the random noise, and
hence have an estimate of the {\it rms} of the noise. Because the
noise power changes with frequency, the errors also become frequency
dependent: the shaded area in a figure analogous to
Fig.~\ref{fig:rskew3s} no longer has constant width.

\section{Discussion}\label{sec:disc}

We have demonstrated that the skewness could be a useful tool for
signal extraction in the presence of realistic overall levels of
foregrounds and noise. It could be, though, that some aspect we have
not accounted for makes the process more difficult. For example, while
the foreground maps we use have {\it rms} fluctuations of the correct
magnitude, they have rather low levels of skewness, perhaps
unrealistically low. If, instead, the foregrounds turn out to be very
skewed, then unless the algorithm to fit out foregrounds is unbiased,
this could propagate into the fitting residuals and drown out the
CS. In this case, however, the characteristic pattern
in the skewness as a function of redshift -- a dip followed by a peak
-- may allow the signal to be picked out even in the presence of
residuals from the foregrounds. This would be exploiting the
smoothness of the foregrounds as a function of frequency once more.

As pointed out by \citet{JEL08}, data constraining the characteristics
of the foregrounds at the relevant scales and frequencies are quite
scarce (though see, for example, \citealt{PEN08}). None the less, the
extrapolations we make from larger scales and higher frequencies may
be pessimistic, if anything. Moreover, the structured noise appears to
be at least as influential as the foreground fitting residuals in
limiting the sensitivity of our signal extraction using the
skewness. For other statistics, great care may be required to model
and remove the foregrounds to high accuracy \citep*{MOR06}. Dealing
with polarized foregrounds is also a concern, which will be approached
in future work.

Clearly, it is also desirable to model the CS itself
accurately, especially when using higher order statistics. A good
extraction scheme should work for a wide range of reionization
scenarios, and ideally should be able to distinguish between them. We
have therefore tested our scheme with three detailed models in which
the ionization history is quite different. Though in each case we see
the skewness become relatively small before rising to large values --
behaviour which may be generic -- a more extended period of
reionization stretches out these features. It is possible, however,
that more exotic sources which we have neglected would cause different
behaviour. If, for example, decaying dark matter makes a significant
contribution to heating the intergalactic medium before reionization
or during its early stages, we can expect this heating to be uneven,
the rate of energy deposition depending on the square of the
density. Then we can no longer assume that the hydrogen spin
temperature, $T_\mathrm{s}$, is much larger than the CMB temperature,
$T_\mathrm{CMB}$, everywhere. We would have to multiply
Equation~\ref{eqn:dtb} by a position-dependent factor
$(T_\mathrm{s}-T_\mathrm{CMB})/T_\mathrm{s}$, which could result in
non-zero skewness even if the neutral fraction is approximately unity
everywhere.

The main limitation of these simulations when it comes to testing our
extraction scheme, however, is their size. To generate maps with the
area of one LOFAR EoR window we must tile our datacubes in the image
plane. In some cases this may be unrealistic: when the size of
individual ionized bubbles becomes comparable to the size of the
simulation box, a slice through the box can no longer be considered to
be a representative slice of the Universe. We have argued that this
may not be important for one-point statistics (and if anything, having
a larger number of independent volumes contributing to each image
would improve our signal to noise ratio). Firstly, each pixel is in
any case affected differently by foregrounds and noise which are much
larger than the CS. Secondly, nearby frequency slices
are at a similar stage of reionization but may otherwise be
sufficiently weakly correlated that smoothing along the frequency
direction after extraction can help recover a clearer trend. Spatial
statistics will clearly be more seriously affected by tiling. Note
that there are also periodic repetitions in the frequency (redshift)
direction in the simulated CS. This can be seen in
the high redshift portion of the curve corresponding to the evolution
of $\langle\delta T_\mathrm{b}\rangle$ in the f250C simulation in
Fig.~\ref{fig:skewandmean}. The onset of reionization appears to break
this periodicity somewhat: for example, the mean ionized fraction can
change significantly between two redshifts separated by a comoving
radial distance corresponding to the size of the simulation box. It
therefore seems to be no obstacle to robustly recovering the overall
trends.

\section{Summary and conclusions}\label{sec:conc}

Many statistics have been put forward to characterize the 21cm
emission from the EoR, the power spectrum probably being the most
frequently studied \citep[][to choose some recent
examples]{BAR08,LID08,PRI08,SET08}. We have suggested that
higher-order statistics may be useful not only to characterize a
CS cube that has been cleaned of foregrounds, noise
and instrumental effects, but also to extract the signature of
reionization from these corrupting influences in the first place. The
skewness of the one-point distribution of brightness temperature,
measured as a function of observed frequency (or equivalently as a
function of redshift), is one such promising statistic.

The three detailed simulations of reionization which we have studied
show a strong evolution of the skewness with redshift. Some of the
features of this evolution appear to be generic and can be readily
understood: in the early stages of reionization the skewness drops
below that of the underlying density field as the first ionized
bubbles, from which the emission is negligible, are formed. As
reionization progresses, the majority of the volume becomes ionized
and the skewness increases again, becoming very large at low redshift
when the distribution of brightness temperature is peaked at zero,
with a tail extending to large values. In simulation f250C there is a
well defined dip in the skewness with a width $\Delta z\approx 1$. Our
other simulation in which the Universe is reionized entirely by stars
(T-star) shows a more gradual change, with the epoch of reionization
extending throughout the redshift range probed by LOFAR. A third
simulation, T-QSO, in which QSOs reionize the Universe, shows an
intermediate behaviour.

By combining these simulations with models of the foregrounds, noise
and instrumental response, we have generated datacubes which are
intended to simulate the output of the LOFAR EoR experiment. We have
studied two cases: firstly, one in which we smooth the foregrounds and signal
to the resolution of the telescope using a Gaussian kernel, then add
uncorrelated Gaussian noise; secondly, one in which we degrade the
foregrounds and noise to the resolution of the telescope using a
realistic PSF, and add noise which is uncorrelated in the Fourier
plane rather than the image plane, producing what we refer to as
`dirty' images. In the former case, we can see the
signature of reionization in the skewness by fitting out the
foregrounds to obtain residual images, and then denoising these images
with a simple smoothing operation. The skewness in these images as a
function of redshift shows significant evidence of reionization. The
result is quite robust to the details of the foreground fitting and
the smoothing. Under- or over-fitting the foregrounds affects the
recovered skewness less severely than the recovered
variance. Extracting a signal from the dirty cubes requires a more
sophisticated denoising scheme. In an optimistic scenario where the
correlation matrices of the original signal and of the noise are
known, we can again recover the evolution of the skewness quite
cleanly using Wiener deconvolution.

We have touched upon some areas for improvement: simulations which
remain realistic but extend to larger scales and exhibit an even
greater range of reionization histories; taking into account the
polarization of the foregrounds and the instrumental response, and
incorporating new observational constraints as they arrive; testing
the minimal assumptions we must make about the signal in order for our
extraction scheme to work, for example whether a poor estimate of the
correlation matrix of the CS seriously affects the
extracted skewness; and studying a wider range of statistics beyond
the variance and power spectrum. All of these will be areas for future
work. Even at this stage, however, our results justify some optimism
that the new generation of radio telescopes can detect the signature
of reionization using higher-order statistics.

\section*{Acknowledgments}

GH is supported by a grant from the Netherlands Organisation for
Scientific Research (NWO). As LOFAR members, the authors are partially
funded by the European Union, European Regional Development Fund, and
by `Samenwerkingsverband Noord-Nederland', EZ/KOMPAS. GM and II
acknowledge that this study was supported in part by Swiss National
Science Foundation grant 200021-116696/1 and Swedish Research Council
grant 60336701.

\bibliography{allbib}
\end{document}